\documentclass{amsart}
\usepackage{graphicx}
\usepackage{amssymb}
\usepackage{xcolor}

\usepackage[a4paper, total={6in, 8in}]{geometry}
\usepackage{setspace}

\usepackage{lineno}

\usepackage[superscript]{cite}

\usepackage{mhchem}
\usepackage{amsaddr}

\newcommand{\jgr}{Journal of Geophysical Research}
\newcommand{\icarus}{Icarus}

\author{Sean Jordan*$^1$}
\author{Oliver Shorttle$^{1,2}$}
\author{Paul B. Rimmer$^{2,3,4}$}
\address{$^1$Institute of Astronomy, University of Cambridge, Madingley Rd, Cambridge CB3 0HA, United Kingdom}
\address{$^2$Department of Earth Sciences, University of Cambridge, Downing St, Cambridge CB2 3EQ, United Kingdom}
\address{$^3$Cavendish Laboratory, University of Cambridge, JJ Thomson Ave, Cambridge CB3 0HE, United Kingdom}
\address{$^4$MRC Laboratory of Molecular Biology, Francis Crick Ave, Cambridge CB2 0QH, United Kingdom}
\address{*corresponding email: saj49@ast.cam.ac.uk}

\title{Proposed energy-metabolisms cannot explain the atmospheric chemistry of Venus}

\begin{document}

\maketitle

\begin{abstract}

Life in the clouds of Venus, if present in sufficiently high abundance, must be affecting the atmospheric chemistry. It has been proposed that abundant Venusian life could obtain energy from its environment using three possible sulfur energy-metabolisms. These metabolisms raise the possibility of Venus's enigmatic cloud-layer \ce{SO2}-depletion being caused by life. We here couple each proposed energy-metabolism to a photochemical-kinetics code and self-consistently predict the composition of Venus's atmosphere under the scenario that life produces the observed \ce{SO2}-depletion. Using this photo-bio-chemical kinetics code, we show that all three metabolisms can produce \ce{SO2}-depletions, but do so by violating other observational constraints on Venus's atmospheric chemistry. We calculate the maximum possible biomass density of sulfur-metabolising life in the clouds, before violating observational constraints, to be $\sim10^{-5}\,-\,10^{-3}\,{\rm mg\,m^{-3}}$. The methods employed are equally applicable to aerial biospheres on Venus-like exoplanets, planets that are optimally poised for atmospheric characterisation in the near future.

\end{abstract}

\label{sec:intro}

\section*{Introduction} 

Sulfur dioxide, \ce{SO2}, is observed to be depleted in the cloud layer of Venus by orders of magnitude compared to its below-cloud abundance \cite{Vandaele2017}. Gas-phase sulfur chemistry at the cloud top is dominated by the photochemical formation of sulfuric acid, where one photon of light, one \ce{H2O} molecule, and up to two \ce{SO2} molecules, are consumed in the production of one \ce{H2SO4} molecule, which then condenses and forms the bulk of the mass loading in the clouds. According to this reaction scheme, the maximum depletion of \ce{SO2} that could possibly be achieved is within a factor of $2$ of the \ce{H2O} mixing ratio, however this is inconsistent with the observations of $\sim 150\,{\rm ppm}$ below-cloud \ce{SO2} and $\sim 10\,{\rm ppb}$ above-cloud \ce{SO2}, yet only $\sim 30\,{\rm ppm}$ below-cloud \ce{H2O} -- a factor of five lower than the \ce{SO2} mixing ratio. Thus, Venus's observed \ce{SO2} depletion represents a key gap in our understanding of the planet's atmosphere \cite{YungDeMore1982, Parkinson2015, Marcq2018}, for which no full-atmosphere model can self-consistently predict via exclusively gas-phase chemistry \cite{BiersonZhang2020}. This has led to the conclusion that there must be some currently unknown chemical pathway responsible for the depletion of \ce{SO2} in the middle atmosphere. One abiotic solution has been proposed which hypothesises that the delivery of hydroxide salts to the cloud layer in the form of mineral dust can enable reduction of \ce{SO2} within the cloud droplets to reproduce the observed trend \cite{Rimmer2021}. It is not yet known whether this in-droplet mineral chemistry is occurring in the clouds of Venus, or whether the delivery of mineral dust to the cloud droplets is sufficient for this mechanism to produce the observed \ce{SO2} depletion. This raises the possibility of an alternative hypothesis: the observed depletion of \ce{SO2} in the clouds of Venus is due to the bio-chemistry of sulfur-metabolising life in an extant aerial biosphere harboured within the cloud layer. 

While the present day surface of Venus is uninhabitable, the lower cloud layer lies in the region of the atmosphere where pressure-temperature conditions are suitable for life (47-57 km altitude). This has led to suggestions of microbial life being a possible explanation for the existence of large and irregularly shaped (`mode 3') aerosol particles in the lower clouds and being the identity of the unknown UV-absorber \cite{MorrowitzSagan1967,Grinspoon1997,Limaye2018}. It has recently been suggested that the metabolic activity of microbial life in the clouds could also be the origin of the reduced chemical species \ce{PH3} and \ce{NH3}, which are biotically produced on Earth and have been tentatively observed in the Venusian atmosphere at abundances many orders of magnitude greater than those predicted given their short photochemical lifetimes \cite{Greaves2020a,Seager2020,Bains2020, Mogul2021}. A plausible life cycle for a strictly aerial biosphere in the clouds has been investigated \cite{Seager2020}, and a suite of mission concepts for a future Venus Life Finder Mission have been recently presented to the community \cite{Seager2021}. It it therefore timely to provide novel and quantitative tests of the potential metabolisms of a hypothetical aerial biosphere to better understand the requirements and limits of life in the clouds, if it exists. Given that the terrestrial biosphere has had an inexorable effect on the Earth's atmospheric chemistry throughout geological time, investigating Venusian biochemistries self-consistently with the atmospheric chemistry is crucially important to assessing the viability of the aerial life-hypotheses.

It has historically been proposed that hypothetical Venusian life could have originated at the planet's surface or been seeded from Earth, if Venus was ever habitable in its past. This surface-dwelling life would have subsequently migrated to the temperate region of the cloud layer as Venus underwent runaway greenhouse warming, residing in the clouds today as an aerial biosphere \cite{MorrowitzSagan1967, Cockell1999, Grinspoon1997}. Runaway greenhouse warming is instigated by an increase in the surface temperature of a planet which causes an increase in the evaporation of surface water to the atmosphere. Since \ce{H2O} is a strong greenhouse gas, the increase in atmospheric \ce{H2O} further exacerbates the warming effect, creating a positive climate feedback. Atmospheric \ce{H2O} is eventually lost via photo-dissociation followed by hydrogen loss to space, and the record of the past water content is left imprinted in the atmospheric deuterium to hydrogen ratio (D/H). The enhanced D/H ratio measured in the atmosphere of Venus provides evidence of past water loss \cite{deBergh1991}, however it is unclear whether this water was ever present as liquid at the surface or always as gas in a hot steam atmosphere. If liquid water was present at the surface of Venus in its past then the planet's slow rotation rate would have enabled a stabilising cloud-climate feedback mechanism to be established \cite{WayDelGenio2020}, maintaining habitable surface conditions despite the increasing Solar luminosity over time. This phase of habitability could have lasted for up to $\sim 900$ Myrs if Venus has always been in a stagnant lid tectonic regime \cite{Honing2021}, or if early Venus had Earth-like plate tectonics then this habitable period could have lasted for up to $4$ Gyrs, until runaway greenhouse was instigated by voluminous magmatism \cite{WayDelGenio2020}. In contrast, other modelling has suggested that, due to its proximity to the Sun, Venus could never have cooled sufficiently for liquid water to condense at the surface following its magma ocean phase, and thus a stabilising cloud feedback could never have been established \cite{Turbet2021}. In this case, life in the clouds would likely require an aerial origin of life scenario within cloud droplets, utilising very different biochemistry to terran life.

Regardless of the proposed origins or adaptations of hypothetical life on Venus, the requirement of life to obtain energy and materials from its environment is universal. A stable biosphere must consume metabolic feedstock molecules and release metabolic products to liberate energy regardless of the particulars of the biochemical machinery. A Venusian aerial biosphere has been suggested to be dependent on a sulfur-based energy-metabolism\cite{SchulzeMakuch2004,SchulzeMakuchIrwin2006} for three reasons: 1) the chemical energy liberated via currently proposed sulfur-based metabolisms \cite{SchulzeMakuch2004,SchulzeMakuchIrwin2006} is sufficient to meet the energy requirements of a cell \cite{LaRoweAmend2015a}; 2) the relatively high abundance of sulfur species (e.g., \ce{SO2, OCS, H2S}) in the Venusian atmosphere means these are the abundant feedstock molecules that could support a global biosphere; and, 3) terrestrial acidophillic organisms provide a precedent for life in an acid environment which uses sulfur-metabolism as a primary source of energy (e.g., acidithiobacillus ferooxidans). Further support for sulfur-metabolising life in Venus's clouds comes from an extensive analysis into the spectral absorptivities of proteins and cofactors used by terrestrial sulfur-metabolising microbes. Significant overlap was found between the spectral appearance of these molecules and the UV features of Venus's spectrum, currently attributed to the unknown absorber \cite{Limaye2018}. 

Three sulfur-based energy-metabolisms for putative Venusian microbes have been proposed by Schulze-Makuch (2004) and Schulze-Makuch \& Irwin (2006) \cite{SchulzeMakuch2004,SchulzeMakuchIrwin2006}, each of which utilise the relatively abundant sulfur species in the atmosphere and can, in principle, induce the depletion of \ce{SO2} as a result of their respective biochemical effects on the atmospheric chemistry. One proposed metabolsim, `A', is a primitive anoxygenic photosystem used by microbes on Earth, oxidising hydrogen sulfide (\ce{H2S}) to elemental sulfur (\ce{S}) meanwhile fixing carbon from atmospheric \ce{CO2} into organics \cite{SchulzeMakuch2004}:
\begin{linenomath*}
\begin{align}
2\,\ce{H2S}  + \ce{CO2} &\ce{->[light]} (\ce{CH2O}) + \ce{H2O} + \ce{S2}, \label{eq:photosystem}
\end{align}
\end{linenomath*}
where (\ce{CH2O}) represents carbon locked in organic molecules (e.g., glucose). We couple this photosystem to reaction \ref{eq:sulfur_respiration}, where the organics produced by the anabolism of reaction \ref{eq:photosystem} are consumed in catabolism to liberate chemical energy:
\begin{linenomath*}
\begin{align}
\ce{SO2}  + (\ce{CH2O}) &\rightarrow \ce{CO2} + \ce{H2O} + \ce{S}. \label{eq:sulfur_respiration}
\end{align}
\end{linenomath*}
We have constructed reaction \ref{eq:sulfur_respiration} as a hypothetical variant of terrestrial oxygenic respiration with \ce{O2} replaced by \ce{SO2}. The widespread use of \ce{O2} for respiration by life on Earth is believed to be an adaptation of anoxygenic life following oxygenation of the Earth's atmosphere to contain abundant \ce{O2}\cite{Soo2017}. In contrast, \ce{O2} is not observed in the upper atmosphere of Venus leading to an inferred upper limit on its abundance of $<2.8\,{\rm ppm}$ \cite{Marcq2018}. One measurement of \ce{O2} within the cloud layer suggests an abundance of $\sim70\,{\rm ppm}$\cite{Oyama1979} however this detection remains controversial, and is discrepant with the results of modelling\cite{Rimmer2021} and the subsequent non-detection of \ce{O2}\cite{Mills1999}. Taking the widely accepted view that there is not substantial \ce{O2} in the atmosphere of Venus, we assume that hypothetical Venusian life would have evolved to utilise the abundant \ce{SO2} in the atmosphere in an analogous fashion to terrestrial oxygenic life's respiration of atmospheric \ce{O2}. The net metabolic reaction for metabolism A is therefore:
\begin{linenomath*}
\begin{align}
\text{Metabolism A (net):}\quad 2\,\ce{H2S} + \ce{SO2} &\ce{->[light]} 2\,\ce{H2O}  + \ce{S2} + \ce{S}.
\label{eq:phot_reaction}
\end{align}
\end{linenomath*}
Reaction \ref{eq:phot_reaction} is expressed in it's simplest stoichiometric ratio. In reality life is more likely to be be performing higher multiples of reaction \ref{eq:phot_reaction}, constructing and burning longer chain organic molecules, and releasing longer sulfur allotropes, rather than \ce{S} and \ce{S2}. We use this representation of the metabolism for modelling simplicity when coupling biochemistry to the atmospheric chemistry, and this does not impact any of the results that we will present.

Metabolism B is a chemoautotrophic pathway proposed in the literature for hypothetical Venusian life, that exploits the redox disequilibrium between \ce{CO}, \ce{H2} and \ce{SO2} in the atmosphere \cite{SchulzeMakuchIrwin2006}:
\begin{linenomath*}
\begin{align}
\text{Metabolism B:}\quad \ce{H2}  + 2\,\ce{CO} + \ce{SO2} &\rightarrow 2\,\ce{CO2} + \ce{H2S}.
\label{eq:chem_1_reaction}
\end{align}
\end{linenomath*}
Metabolism C is also a chemoautotrophic pathway proposed in the literature for hypothetical Venusian life, and exploits the redox disequilibrium between \ce{CO} and \ce{SO2} in the atmosphere \cite{SchulzeMakuchIrwin2006}:
\begin{linenomath*}
\begin{align}
\text{Metabolism C:}\quad 3\,\ce{CO} + \ce{SO2} &\rightarrow \ce{OCS} + 2\,\ce{CO2}.
\label{eq:chem_2_reaction}
\end{align}
\end{linenomath*}
Both of the proposed chemotrophic metabolisms, B and C, could yield $\sim 240\,{\rm kJ mol^{-1}}$ of energy to a putative Venusian microbe, which would be a sufficient source of chemical energy for life to thrive \cite{LaRoweAmend2015a}, provided the reactant species are in high enough abundance.

We present a framework for the coupling of each proposed energy-metabolisms to a photochemical-kinetic network for modelling planetary atmospheres in 1D \cite{RimmerHelling2016,Rimmer2021}, and self-consistently solve for the atmospheric composition of Venus required to reproduce the observed \ce{SO2}-profile biochemically. We incorporate each sulfur-based metabolism into our reaction network in turn, impose that the reaction rate can only be non-zero between $47-57\,{\rm km}$ altitude (where ambient atmospheric temperatures are between $0 - 100\,{\rm^{\circ}C}$ and aerosols are large enough to accommodate microbial colonies \cite{Seager2020}), and investigate the parameter space of metabolic activity versus availability of metabolic inputs. While plausible metabolic pathways and survival strategies for an extant Venusian biosphere have been suggested in previous studies \cite{Cockell1999,SchulzeMakuch2004,SchulzeMakuchIrwin2006,Limaye2018,Seager2020} none have been investigated self-consistently with the wider atmospheric chemistry. By coupling proposed metabolic requirements with self-consistent models of atmospheric chemistry, we here produce a rigorous test of whether the proposed metabolisms could support a Venusian aerial biosphere. The results of our investigation show that, while proposed metabolic incorporation of gas-phase species can reproduce the sulfur-depletion, in every case the resultant atmospheric chemistry violates separate observational constraints on the lower atmosphere of Venus.

\section*{Results}
\label{sec:results}

\subsection*{Predicting \ce{SO2}-depletion biochemically.} 
\begin{figure}[ht!]
\includegraphics[width=\textwidth]{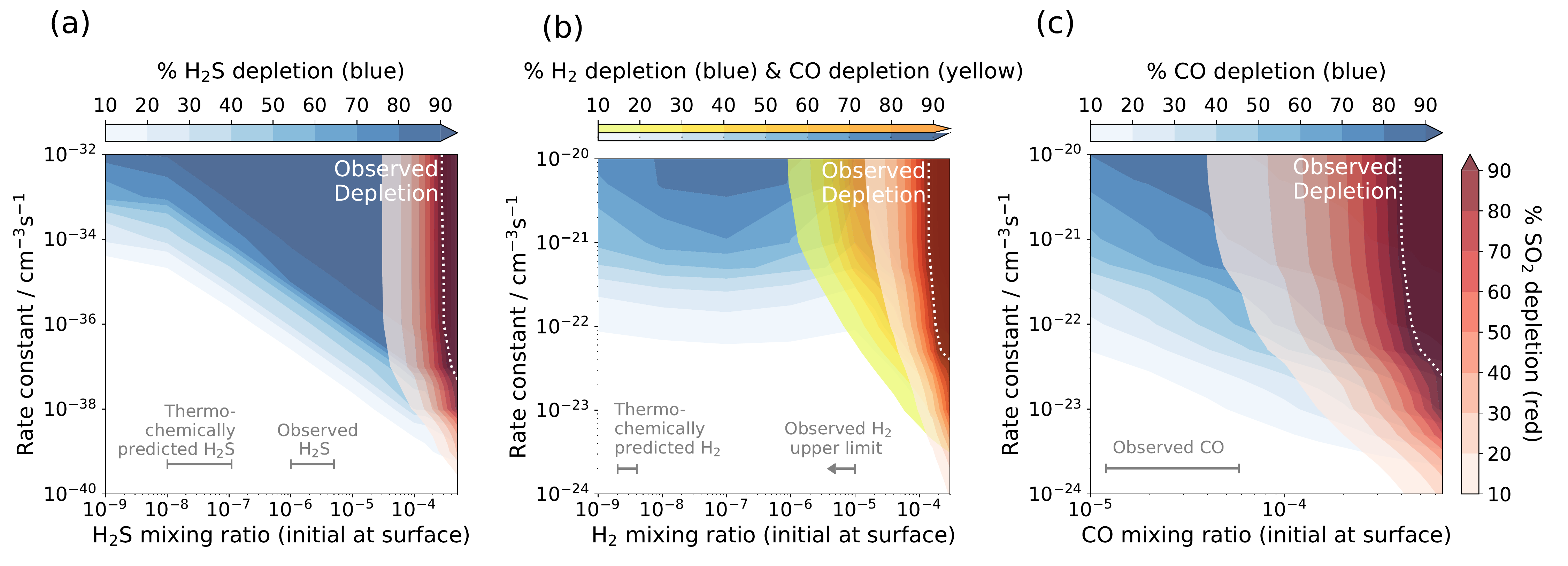}
\caption{\textbf{Depletion contours of metabolic inputs.} Contour plot showing depletion of metabolic reagents within the cloud layer for each metabolic pathway, (a), (b), and (c). Depletion is shown as a function of net metabolic rate (y-axis) and initial surface mixing ratio of the limiting metabolic reagent species (x-axis). The depletion of each species is calculated with respect to the species' column abundance in a reference model run without the metabolism present. The predicted or observed surface mixing ratios of metabolic reagent species and their 1$\sigma$ errors are indicated with grey errorbars \cite{Rimmer2021}, and the metabolic models reproducing \ce{SO2}-depletion are indicated with a white dotted line. \label{fig:contours}}
\end{figure}

Each of the three metabolisms are capable of reproducing the observed \ce{SO2}-depletion biochemically in the middle atmosphere of Venus when coupled to our photochemical-kinetics network. For each metabolism, Figure \ref{fig:contours} shows the depletion of the metabolic reagent species, above $47\,{\rm km}$ altitude, as a percentage of the metabolism-free reference abundance. The figure describes where each metabolic pathway is rate-limited (within the white space) or reagent-limited (within the darkest coloured contour for a given reagent species), with transitional regions between each limiting regime. The resultant \ce{SO2}-depletion is shown with red contours in every panel. We take a surface boundary condition for the initial chemical composition that is consistent with past lower-atmosphere \cite{Kras2007,Kras2013} and full-atmosphere \cite{BiersonZhang2020,Rimmer2021} models as our fiducial initial atmospheric composition. This surface condition corresponds to species that have either been observed or are thermochemically predicted to be present at the base of the Venus atmosphere \cite{Rimmer2021}. For the fiducial atmospheric composition, each metabolism is either rate-limited or reagent limited: metabolism A is limited by the \ce{H2S} abundance; metabolism B by the \ce{H2} abundance; and metabolism C by the \ce{CO} abundance.  Therefore, none of the three metabolisms initially provide a significant destruction pathway for \ce{SO2}. 

To fit the \ce{SO2} observations, each metabolism must be in the \ce{SO2}-limiting regime, and thus the abundances of the other limiting metabolic inputs must be increased from the fiducial (and observationally constrained) surface boundary conditions. Figure \ref{fig:contours} demonstrates how, for each metabolic pathway, the net metabolic rate and atmospheric composition can be tuned to successfully reproduce the observed \ce{SO2} profile (white dotted line inside the \ce{SO2}-limiting region).

\subsection*{Comparing the resultant atmospheric chemistry with observational constraints.}
\begin{figure}[ht!]
\includegraphics[width=\textwidth]{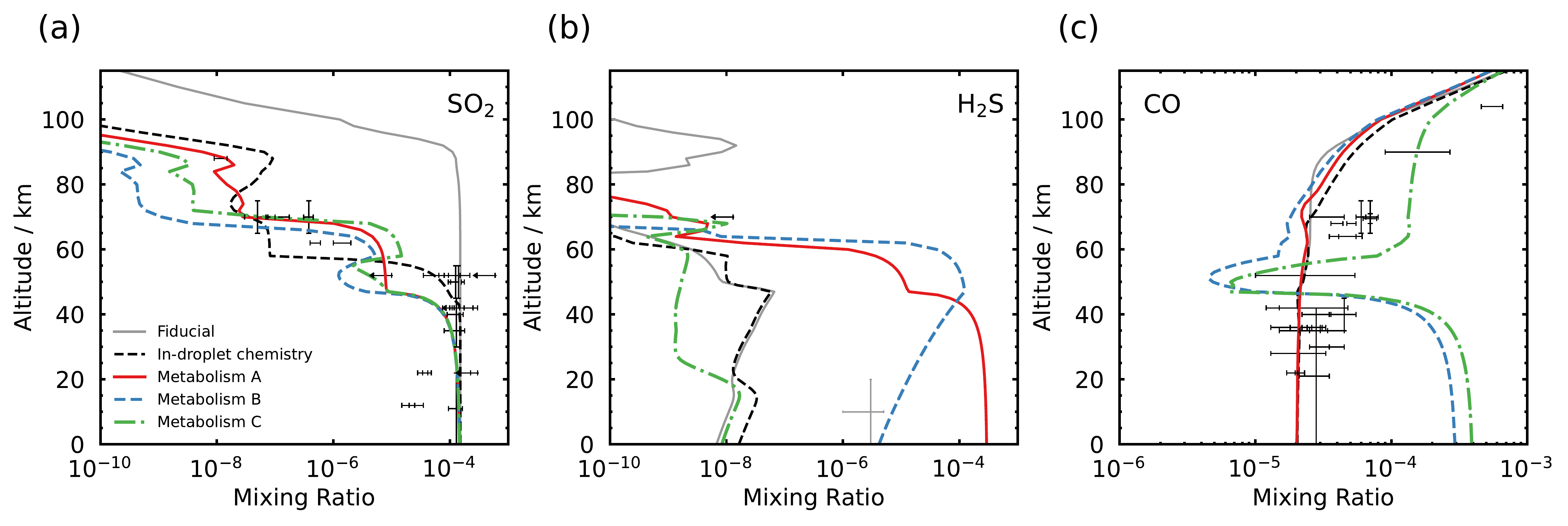}
\caption{\textbf{Atmospheric profiles resulting from biochemical \ce{SO2}-depletion.} Atmospheric profiles of (a) \ce{SO2}, (b) \ce{H2S} and (c) \ce{CO} mixing ratios as a function of altitude for the Rimmer et al., (2021) models (with and without droplet chemistry) \cite{Rimmer2021}, compared with the three proposed metabolic pathways \cite{SchulzeMakuch2004,SchulzeMakuchIrwin2006} under conditions that reproduce the \ce{SO2} observations. Observational data and their 1$\sigma$ errors are shown in black except for the observation of \ce{H2S} below $20\,{\rm km}$ altitude which is shown in grey, as this is possibly a spurious detection \cite{HoffmanHodgesDonahue1980}. \label{fig:profiles}}
\end{figure}

In order to reproduce the observed \ce{SO2}-depletion biochemically, the predicted abundance of metabolic input species for each of the three metabolisms violates a separate observational constraint in the lower atmosphere of Venus. In figure \ref{fig:profiles} we plot the mixing ratios of \ce{SO2}, \ce{H2S} and \ce{CO} as functions of altitude, for the model atmosphere that best fits the \ce{SO2} observations for each metabolic pathway. Alongside, we show the fiducial atmosphere, predicted from thermochemistry, photochemistry and chemical kinetics, which does not correctly predict the \ce{SO2} observations, and the abiotic model atmosphere of Rimmer et al., (2021) which successfully predicts the \ce{SO2} observations by hypothesising aqueous droplet chemistry \cite{Rimmer2021}.  All three proposed metabolisms fit the \ce{SO2}-depletion with a similar accuracy to the droplet chemistry hypothesis \cite{Rimmer2021}, with metabolism A providing the closest fit of any hypothesis to the middle-atmosphere observations (Figure \ref{fig:profiles}). However, this comes at the cost of the other metabolic reagent species not then fitting their below-cloud observational constraints well.

Metabolisms B and C each violate well-constrained lower-atmosphere observations of \ce{CO}, requiring $300\,{\rm ppm}$ and $400\,{\rm ppm}$ of \ce{CO} at the surface respectively, approximately $\sim 15 \times$ and $\sim 20 \times$ greater than observations indicate. Below-cloud \ce{CO} has been measured in situ by gas chromatography instruments on board the Venera 12 and Pioneer Venus descent probes \cite{Gelman1979,Oyama1980}, by remote detection via IR spectra in a $2.3\,{\rm \mu m}$ atmospheric transparency window by the Venus Express orbiter \cite{Marcq2008,Tsang2009}, and via remote Earth-based detection \cite{Bezard1990,Pollack1993}. All of the data generally agree on deep atmosphere \ce{CO} mixing ratios between $17 - 45\,{\rm ppm}$ and therefore a \ce{CO} mixing ratio as high as that required by metabolisms B and C would be unexpected. 

Metabolism A requires $300\,{\rm ppm}$ of \ce{H2S} at the surface in order for life to be responsible for the observed \ce{SO2}-depletion. \ce{H2S} has been measured at an abundance of $3\,{\rm ppm}$ below the cloud layer via a mass spectrometer instrument on-board the Pioneer Venus descent probe \cite{HoffmanHodgesDonahue1980}, however this reported value is one to two orders of magnitude larger than the currently expected surface mixing ratio of $\sim 10 - 100\,{\rm ppb}$ from thermochemical modelling \cite{Kras2007,Rimmer2021}. The \ce{H2S} abundance otherwise only has an inferred upper limit at the cloud top derived from non-detection by Earth-based platforms \cite{Kras2008}. From Figure \ref{fig:profiles} we find that, even when the \ce{H2S} mixing ratio is on the order of $100{\text{'s}\,\text{ppm}}$ in the cloud layer, \ce{H2S} is so efficiently photolysed by the irradiating Solar flux that the profile always remains consistent with the inferred upper limit at the cloud top (with the \ce{H} atoms liberated forming \ce{H2}, and the sulfur atoms liberated mainly forming sulfur allotropes, \ce{S_x} for $x=1,...,8$). Given the relative lack of observational constraints on the deep atmosphere \ce{H2S} abundance, it remains a remote possibility that \ce{H2S} exists with the required surface mixing ratio for \ce{SO2}-depletion via metabolism A. However, assuming a volcanic source of \ce{SO2} and \ce{H2S} in the Venus atmosphere \cite{Fegley2014}, this would require that the molar ratio of volcanically degassed \ce{H2S/SO2} is $\sim 2$. This is in disagreement with the oxidising nature of the Venus atmosphere, the observation of oxidised sulfur-containing minerals at the Venera 13, Venera 14, and Vega 2 landing sites \cite{Surkov1984,Surkov1986,Fegley2014}, and the inference that the volcanic \ce{H2S} abundance should be lower than that of \ce{SO2} due to the very low water content in the atmosphere \cite{Fegley2014}. 

\section*{Discussion}
\label{sec:discussion}

Our results demonstrate that the three sulfur-based metabolic pathways proposed for Venusian aerial life are capable of reproducing the observed \ce{SO2}-depletion in the cloud layer of Venus, but in each case they require a source of chemical reducing power roughly equal in abundance to below-cloud \ce{SO2}: the atmosphere otherwise does not possess enough reducing power for life to exploit to generate the observed \ce{SO2}-depletion.

It is possible that our understanding of the below-cloud and in-cloud atmosphere is not correct. There is a consensus picture of Venus's atmosphere, however this picture still has several mysteries that models are not able to solve, and certain observational data are discrepant with this picture. Discrepant measurements include in-situ and ground-based detections of \ce{H2, H2O, H2S}, and \ce{O2}, at higher abundances than expected\cite{Oyama1979,Oyama1980,HoffmanHodgesDonahue1980,Mukhin1982}, although these data are not consistent with each other without introducing spatial and temporal variability or significant systematic error. If the discrepant measurements were not made in error, then they must somehow be reconciled with the known atmospheric chemistry and thus reveal potential uncertainties in the consensus picture. However, even if these outliers represented the global atmosphere of Venus, they would still be inconsistent with these metabolic pathways explaining the sulfur-depletion in Venus's atmosphere. If, as seems likely, the standard description of the lower-atmosphere composition of Venus is broadly correct \cite{Rimmer2021}, then we must conclude that either the \ce{SO2}-depletion is due to a different metabolic pathway, or \ce{SO2}-depletion on Venus is not related to life.

If any diagnostic feature of a biosphere utilising the proposed sulfur energy-metabolisms is being imprinted in the atmospheric chemistry of Venus, then the enigmatic cloud-layer \ce{SO2}-depletion is likely to be it. This is because the observed \ce{SO2}-depletion does not yet have a proven abiotic solution, even though the rest of the atmospheric sulfur-chemistry can be explained by abiotic models to within at least an order of magnitude accuracy \cite{Rimmer2021}. It may be that the products of these metabolisms could in principle provide good biosignatures, but only in the case that their presence or abundances are not explained by abiotic sources and sinks. This criteria for metabolic products to be diagnostic biosignatures is not met by the three suggested metabolisms that we investigate here: the presence of \ce{CO2}, \ce{H2O}, \ce{H2S}, \ce{OCS}, and sulfur allotropes in the atmosphere of Venus is already consistent with an abiotic source from volcanic degassing \cite{Fegley2014}; the enhanced abundance of \ce{OCS} and \ce{H2S} produced via metabolisms B and C remain undetectable above the cloud layer due to efficient photochemical destruction by the Solar flux; and any \ce{H2O} produced via metabolism A would very likely be retained by the hypothetical microorganism as part of its survival strategy rather than being released back to the atmosphere, given the environmental stress posed by the relative desiccation of Venus's atmosphere compared to the Earth's atmosphere \cite{Hallsworth2021}. Since there are limited ways to utilize \ce{SO2} as a metabolic input, and since all other reactants would seem to exist in even more limited abundances than those investigated in the study (\ce{CO, OCS, H2S, H2}), this strongly suggests there is no unknown sulfur-metabolism responsible for the observed \ce{SO2}-depletion. 

Alternatively, a Venusian biosphere may exist and not be influencing the atmospheric chemistry in a diagnostic and observable way, in which case we can now place upper limits on the potential biomass density in the cloud layer before observational constraints are explicitly violated for the three proposed metabolisms. In order to calculate this limiting biomass density, an alternative removal mechanism for cloud-layer \ce{SO2} must be incorporated into the model to self-consistently predict the \ce{SO2}-depletion. One alternative biologically-driven mechanism for removing \ce{SO2} has been suggested by Bains et al., ($2021$)\cite{Bains2021}, which hypothesises that life in the clouds, fixing nitrogen from \ce{N2} and using atmospheric \ce{H2O}, can produce \ce{NH3} that reacts with \ce{SO2} to form ammonium sulfites and sulfates, however this has not yet been demonstrated with fully self-consistent chemical modelling. An abiotic route to removing \ce{SO2} in the cloud layer has been suggested by Rimmer et al., ($2021$) by incorporating aqueous droplet chemistry. The droplet chemistry hypothesis posits that an additional source of hydrogen could be delivered to the clouds as hydroxide salts in mineral dust. This would enable aqueous reactions within the cloud droplets to provide a destruction pathway for aqueous \ce{SO2} dissolved in the droplets from the gas-phase. A dust flux to the clouds of $\sim\,16\,{\rm Gt/yr}$ containing $5\,{\rm wt.\%}$ salt would be sufficient to explain the observed \ce{SO2}-depletion. For more details see Rimmer et al., ($2021$) \cite{Rimmer2021}. We here repeat our metabolic analysis, now coupling each metabolism in turn to the abiotic model that incorporates droplet chemistry in order to reproduce the observed \ce{SO2}-depletion self-consistently \cite{Rimmer2021}, and investigate how far we can increase the metabolic activity and abundance of metabolic input species within observational limits. 

\begin{figure}[ht!]
\includegraphics[width=\textwidth]{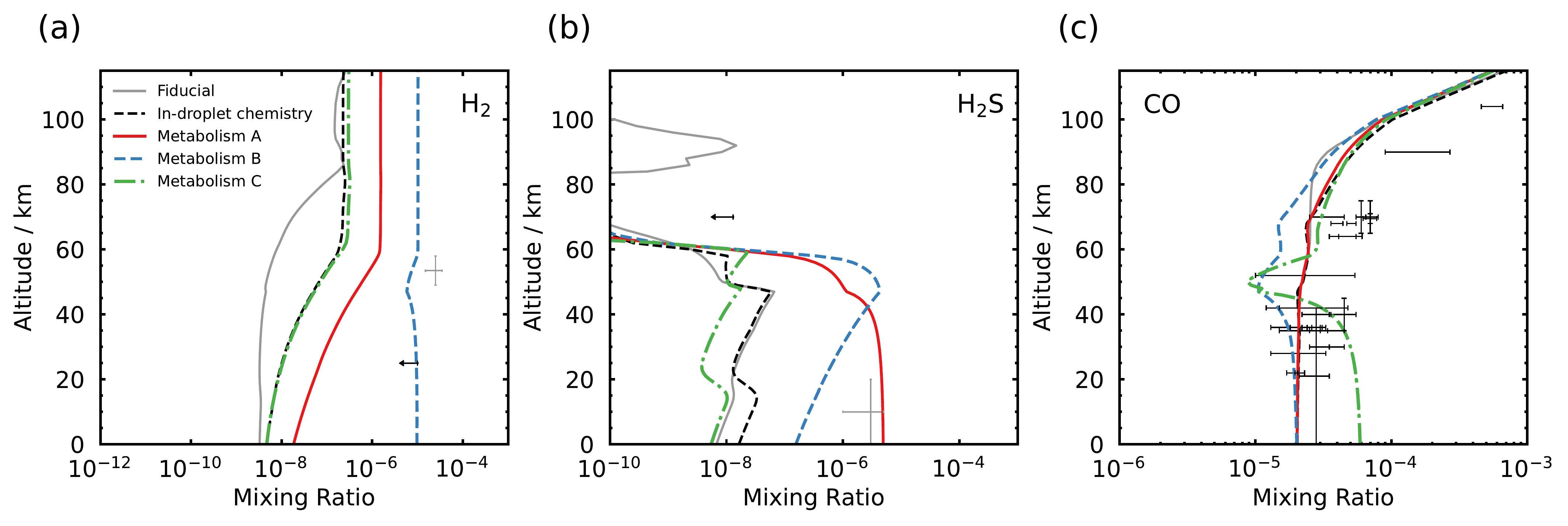}
\caption{\textbf{Atmospheric profiles resulting from the maximum limit of metabolic activity within observational constraints.} Atmospheric profiles of (a) \ce{H2}, (b) \ce{H2S} and (c) \ce{CO} mixing ratios as a function of altitude for the fiducial model and droplet-chemistry (abiotic) model \cite{Rimmer2021}, compared with the three proposed metabolic models at the limit of violating an observational constraint in the atmosphere. Observational data and their 1$\sigma$ errors are shown in black except for the observation of \ce{H2S} below $20\,{\rm km}$ altitude \cite{HoffmanHodgesDonahue1980}, and the observation of \ce{H2} at $50\,{\rm km}$ altitude \cite{Mukhin1982}, which are shown in grey, as these are possibly spurious detections. \label{fig:profiles_limit}}
\end{figure}

The first observational constraints to be exceeded as the biomass density is increased are the constraints on the abundances of metabolic inputs below the clouds. Figure \ref{fig:profiles_limit} shows the three metabolic models with their metabolic input abundances increased as far they can be while remaining just within an observed upper-limit, or within observational error. For metabolism B this corresponds to a $\sim 10\,{\rm ppm}$ upper limit on \ce{H2} below the cloud layer \cite{Donahue1997,Oyama1980}, and for metabolism C the upper end of the estimated error on the largest measured \ce{CO} mixing ratio below the clouds ($\sim 60\,{\rm ppm}$). For metabolism A, assuming that the in situ observation of $3\,{\rm ppm}$ at the surface is correct, the maximum below-cloud \ce{H2S} within observational uncertainty corresponds to $\sim 5\,{\rm ppm}$ \ce{H2S} \cite{HoffmanHodgesDonahue1980}.

\begin{figure}[ht!]
\includegraphics[width=\textwidth]{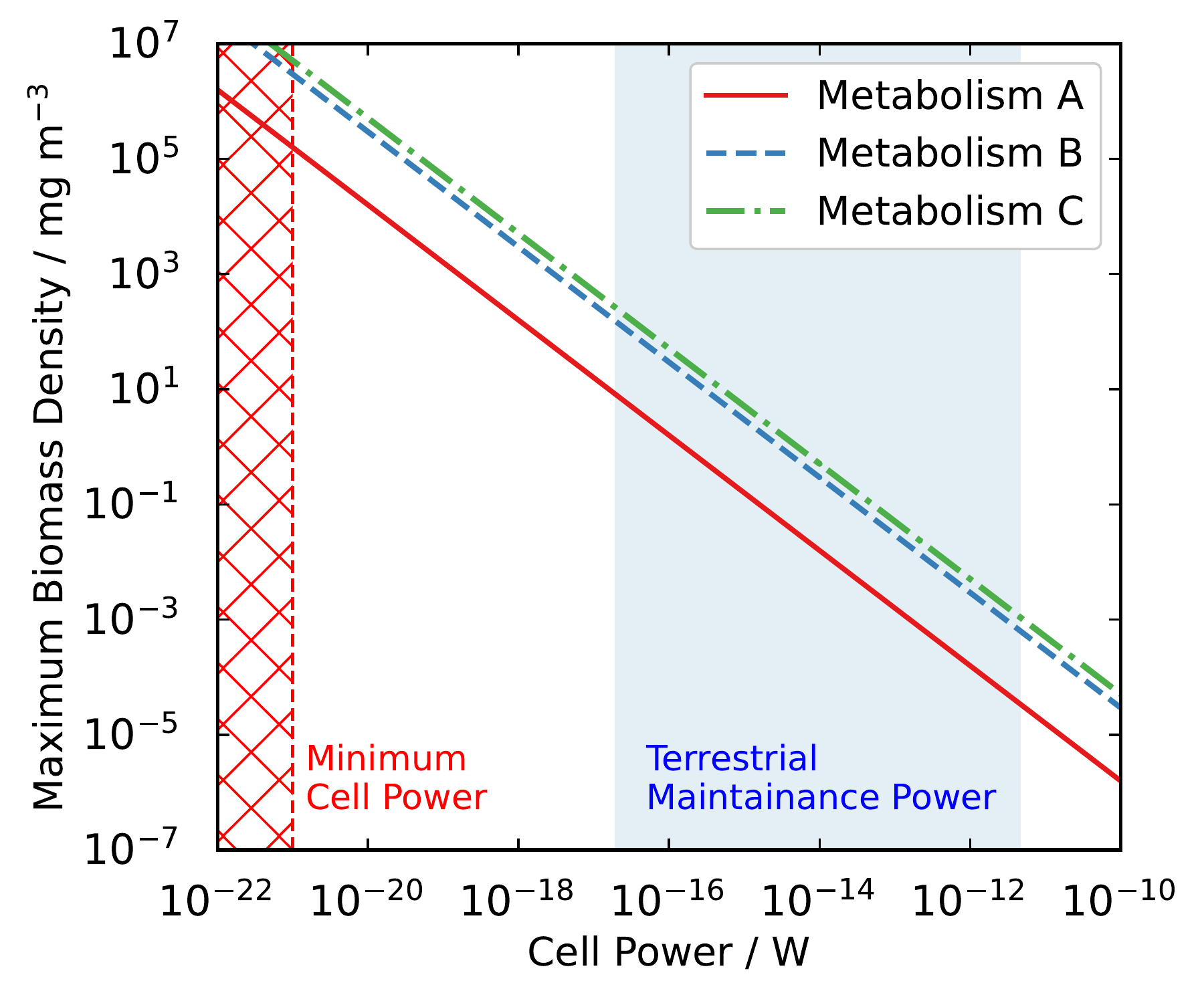}
\caption{\textbf{Maximum biomass density as a function of cell power requirement in the Venusian atmosphere.} Biomass density in the cloud layer for metabolisms A, B, and C, as a function of cell power. The red dashed line indicates the estimated minimum limit on cell power for terrestrial microorganisms \cite{LaRoweAmend2015b} and the red hatch region is the excluded region of the parameter space. The blue shaded region indicates the range of usual cell maintenance power requirements, measured in a laboratory setting, for terrestrial microorganisms \cite{LaRoweAmend2015a}. \label{fig:biomass}}
\end{figure}

We estimate the corresponding biomass density in the cloud layer for the three limiting metabolic models using a simple energetic argument. The reactions of metabolisms B and C are chemotrophic and therefore the $\Delta G_r$ liberated by each metabolic reaction is already known. If we make an assumption about the wavelength of light absorbed by the photosystem of metabolism A then an effective $\Delta G_r$ liberated by the net metabolic reaction can be assumed, give or take some efficiency factor and some numerical factor for the ratio of photons required per metabolic input molecule consumed in the synthesis of long-chain organic molecules (these numerical factors would act to reduce the Gibbs free energy per mole for metabolism A, thus our estimate here is a relevant upper limit). The average volumetric rate of the net metabolic reaction thus provides us with an upper limit on the net energy released via the metabolic activity. We can therefore make an estimate of the maximum biomass density as a function of cell power, under the assumption that cells are $\sim 0.5\,{\rm \mu m}$ in radius \cite{Limaye2018} and the density of cellular material is of the order $\sim 1\,{\rm g\,cm^{-3}}$. The results are shown in figure \ref{fig:biomass}, and further details of this calculation can be found in the Methods section.

The possible biomass range in our model is critically dependent on the power requirements of a single cell. We can first obtain an absolute upper bound on the possible biomass density in the Venusian cloud layer for each proposed energy-metabolism, by adopting that $P_{\rm cell} \approx P_{\rm cell,\, min}$. The red dashed line in Figure \ref{fig:biomass} shows the estimated minimum cell-power, $P_{\rm cell,\, min} \sim 10^{-21}\,{\rm W}$ for terrestrial microbial life \cite{LaRoweAmend2015b}. The biomass density estimates corresponding to this $P_{\rm cell,\, min}$ are $2 \times 10^5\,{\rm mg\,m^{-3}}$ for metabolism A, $3 \times 10^6\,{\rm mg\,m^{-3}}$ for metabolism B, and $5 \times 10^6\,{\rm mg\,m^{-3}}$ for metabolism C. These estimates are inconsistent with the mass loading estimate of $\sim 0.1 - 100\,{\rm mg\,m^{-3}}$ for the Venusian lower-cloud droplets themselves \cite{Limaye2018}, which is problematic as these droplets would host the microbial life in the hypothesised life-cycle \cite{Seager2020}. This treatment must therefore be a large overestimate of the maximum possible biomass because of the minimum power assumption we have made: Life in Venus's clouds would inevitably have far greater power requirements than the minimum estimated power requirements of terrestrial life.

More realistic values for the power requirements of terrestrial microorganisms are those of cell maintenance power -- the power per cell required to prevent population decay -- estimated in a laboratory setting to be between $0.019 - 4700 \times 10^{-15} \, {\rm J\,s^{-1}\,cell^{-1}}$ (indicated with a blue shaded region in Figure \ref{fig:biomass}) \cite{LaRoweAmend2015a}. Using these values, we calculate a range of biomass estimates between $\sim 3 \times 10^{-5} - 8\,{\rm mg\,m^{-3}}$ for metabolism A, $\sim 6 \times 10^{-4} - 160\,{\rm mg\,m^{-3}}$ for metabolism B, or $\sim 1 \times 10^{-3} - 270\,{\rm mg\,m^{-3}}$ for metabolism C. The majority of values in these ranges of maximum biomass density would be consistent with the mass loading in the Venusian cloud layer, and the values at the upper end of the ranges are comparable to the past estimate of $14\,{\rm mg\,m^{-3}}$ \cite{Limaye2018}, and to the estimated biomass density of the Earth's own aerial biosphere, $44\,{\rm mg\,m^{-3}}$ \cite{Limaye2018}.

We can instead express this result in terms of the maximum cell concentration within the aerosol substrate by considering the number density of cells in the cloud layer and the aerosol mass loading of the Venusian lower-cloud droplets (see the Methods section for details). For our maximum biomass density ranging between order $10^{-5} - 10^2\,{\rm mg\,m^{-3}}$, and aerosol mass loading estimates ranging from $0.1 - 100\,{\rm mg\,m^{-3}}$\cite{Limaye2018}, we obtain an estimated range of values for the maximum possible cell concentration between $10^{5} - 10^{15}\,{\rm cells\,mL^{-1}}$. Comparing to the measured values of cell concentrations of ${10^3 - 10^5\,{\rm cells\,mL^{-1}}}$ from hyperarid environments on Earth such as the Atacama desert\cite{Connon2007}, we find that the lower end of the maximum cell concentration range for a Venusian aerial biosphere intersects the uppermost value measured in hyperarid terran environments, and otherwise all other values of maximum cell concentration in the range exceed cell concentrations observed in hyperarid terran environments.  A Venusian biosphere more abundant than a terrestrial biosphere, even in a hyperarid environment, would be unexpected given extreme stresses in the Venusian environment: to loss of biomass out of the cloud layer; and, to desiccation in the sulfuric acid of the cloud droplets. We note, however, that hyperarid terran environments consist of dry soils whereas the hyperarid environment of the Venusian cloud layer consists of liquid sulfuric acid and thus there is no true analog environment on Earth that can be used as a close comparator to that of Venus.

The vast chemical challenges that the acidic environment of Venus would pose to an aerial biosphere would likely result in maintenance power requirements that are larger than any encountered on Earth, which would favour biomass density and cell concentration values below even the lower limit of our estimated biomass ranges (i.e., $\lesssim 3\times 10^{-5},\, 6\times10^{-4}$, and $1\times10^{-3}\,{\rm mg\,m^{-3}}$ for metabolism A, B, and C, respectively). We note also that these estimates have not factored in limitations on biomass from metabolic processes that are not related to energy-capture but are still necessary for life: fixation of nutrients from the atmosphere for growth, for example the energetically costly process of nitrogen fixation from atmospheric \ce{N2}, and the incorporation and availability of metals for use in enzymes to catalyze reactions. This may pose a problem for the life hypothesis proposed by Bains et al., ($2021$), which relies on life fixing nitrogen from atmospheric \ce{N2} in a non-energy-capture metabolism to explain the \ce{SO2}-depletion biochemically\cite{Bains2021}. Given these additional challenges life faces, we emphasise again that the analysis we present here is the most optimistic scenario possible for the hypothetical life that we have considered, in which the only limit imposed on a hypothetical biosphere is the availability of metabolic inputs required to capture chemical energy. These arguments suggest a very high cell power requirement should be favoured for hypothetical Venusian aerial life, with the consequence that any such sulfur-metabolising biosphere would then have to be extremely low in biomass density, at least four orders of magnitude lower than the terrestrial aerial biosphere, to remain consistent with atmospheric observations.

We can compare the above estimated biomass densities to the abundance of matter required to produce the unexplained spectral properties observed at UV wavelengths in the cloud layer, which are attributed to an `unknown UV-absorber'. It has been suggested that the unknown UV-absorber may be explained by the presence of sulfur-metabolising microorganisms harboured in the clouds, with significant overlap found between the spectral appearance of various proteins and photosynthetic pigments used by some sulfur-metabolising microorganisms on Earth \cite{Limaye2018}. Alternatively, one abiotic contender to explain the UV-absorber is suggested to be a $1\,\%$ solution of \ce{FeCl3} dissolved within $75\,\%$ \ce{H2SO4} cloud droplets \cite{Zasova1981,Kras2017}. Performing a comparison of our biomass density estimates to this abiotic suggestion, $1\,\%$ of the mass-loading of the lower clouds is equal to $10^{-3} - 1\,{\rm mg\,m^{-3}}$ \cite{Limaye2018}. If the maximum biomass density we have calculated above is indeed close to or less than the lower end of our estimated range of values (i.e., large cell power requirements with respect to known terrestrial values), then, assuming the biotic matter can produce similar strength spectral properties as the suggested \ce{FeCl3} molecules, only metabolism C could possibly intersect the lower limit of the absorber's required mass-loading range ($10^{-3}\,{\rm mg\,m^{-3}}$) and metabolisms A and B fall short by at least one order of magnitude. However, if the biological pigments can produce much stronger absorption than simple salts then the potential biomass density we have constrained in the Venusian cloud layer may possibly still be a viable explanation to the unknown UV-absorber. Future studies would be needed to compare the standard absorbance of such proposed biological pigments to the observed spectral properties of the cloud aerosol.

One final consideration that could be made is that of the ecological stoichiometry of a combined ecosystem, featuring a combination of metabolisms in a steady state where the fluxes of metabolic products and inputs consumed are equally balanced. For any significant biosphere an ecosystem is needed and nutrient recycling is critical. Given that the terrestrial biosphere would collapse within a short time period in the absence of nutrient recycling between different species in a diverse ecosystem, it is also unlikely that one species of organism could exist by itself in the cloud layer of Venus, with the speculative exception perhaps of an autotrophic species using photosynthesis and only maintaining a very low biomass, such as that described by metabolism A. We have demonstrated that each of the metabolic hypotheses investigated here can be individually ruled out on the basis of violating lower atmosphere constraints on the required metabolic input abundances, and we will now demonstrate how our results also show that a multi-metabolism ecosystem recycling the required chemical species exclusively within the cloud layer cannot explain the sulfur depletion in Venus's clouds. 

If we were to take, for example, the net effect of metabolism A and B together then the flux of \ce{H2S} consumed by life with metabolism A, could be produced at the same rate by life utilising metabolism B. This results in a net reaction:
\begin{linenomath*}
\begin{align}
2\,\ce{H2} + 3\,\ce{SO2} + 4\,\ce{CO} &\rightarrow 2\,\ce{H2O} + 4\,\ce{CO2} + \ce{S2} + \ce{S} 
\label{eq:ecosystem}
\end{align}
\end{linenomath*}
Even if arguments could be posed for the efficient production of \ce{CO} from the abundant \ce{CO2} in the atmosphere via photolysis or perhaps another metabolism in a hypothetical ecosystem, the reaction inevitably still requires input of \ce{H2} to the atmosphere, inconsistent with observations. It has been suggested that \ce{H2} could be replenished in the atmosphere via hydration reactions of ferrous iron oxide minerals at the surface \cite{Watson1984}. Even if there were sufficient \ce{H2O} in the atmosphere to enable this, the resulting below-cloud \ce{H2} mixing ratio would violate the observational upper limit of $<10\,{\rm ppm}$ \cite{Oyama1980}. Production of \ce{H2} via hydration of ferrous iron oxide mineral dust within the cloud layer exclusively would not be possible because this mineral dust would be dissolved in the sulfuric acid cloud droplets and thus any hydrogen in the reaction would exist as aqueous \ce{H+} ions. Mineral dust therefore cannot provide a source of \ce{H2} localised within the cloud layer as a potential metabolic input and \ce{H2} could only be replenished by surface-atmosphere processes. The observational upper limit on \ce{H2} abundance below the clouds therefore poses a strict upper limit on the productivity of any biosphere recycling nutrients in ecosystem interactions. In the absence of ecosystem interactions, the requirement for metabolic inputs to be replenished by surface-atmosphere processes extends to all chemical species and therefore below-cloud observations of any limiting input species poses a strict upper limit on the productivity of the biosphere. 

Ultimately there simply is not enough reducing power in the lower atmosphere of Venus for the suggested forms of life to exploit in order to deplete \ce{SO2} and thus be of sufficient biomass to influence the global atmospheric chemistry. We have demonstrated this result for life using each of the energy-capture metabolisms proposed in the literature so far, however our results extend generally to: a) chemotrophs exploiting energy gradients from gaseous atmospheric species; or b) phototrophs utilising gaseous atmospheric species as electron donors. The exception to this conclusion is phototrophic life that captures light energy not for \ce{CO2} reduction but to support general cellular demands, in which case the potential biomass density in the clouds cannot be constrained via the availability of metabolic inputs in the atmosphere, and future studies are needed to constrain this potential biomass via the available light flux. There may also still plausibly be a low mass biosphere present, the only observable effect of which is to release relatively small amounts of trace metabolic gases. The recent detections of phosphine gas in the atmosphere of Venus may be consistent with this, in the absence of any abiotic phosphine production pathway \cite{Greaves2020a,Greaves2020b,Greaves2020c,Greaves2021,Bains2020}.

Upcoming missions to Venus may be able to determine whether or not the cloud layer hosts a high mass biosphere using a currently unidentified metabolism, a low mass biosphere producing only trace metabolic products, or no biosphere at all. However, the prospect remains that other, possibly wetter, Venus-like exoplanets could host a habitable niche in the temperate region of their atmospheres. Aerial biospheres in general therefore have significant implications on the number and observability of potentially habitable planets beyond the Solar System. The hypothetical `habitable zone' around a star has resulted in the search for Earth-sized exoplanets within the range of orbital distances for which an Earth-like atmosphere could permit condensation of liquid water at the surface \cite{Kasting1993}. Interior to the habitable zone is the region where surface temperatures are too high for water to condense at the surface, and this region is more likely to be populated by Venus-like exoplanets \cite{Kane2014}. For Venus-like exoplanets with permanent cloud cover in the temperate region of the atmosphere, the possibility of an aerial biosphere extends the inner edge of the habitable zone much closer to the host star, where orbital periods are shorter and transiting planetary signals have enhanced signal to noise ratios, compared to planets within the surface-water habitable zone. The presence of a biosphere at high altitude also means that biologically induced features in the atmospheric chemistry will be imprinted in the observationally accessible region of the atmosphere, compared to surface life obscured by thick cloud cover. It has been demonstrated that the metabolic inputs \ce{H2S} (metabolism A) and \ce{SO2} (metabolisms A, B, and C), and metabolic products \ce{OCS} (metabolism C) and \ce{H2S} (metabolism B), will not be depleted photochemically in the observable region of a Venus-like planet's atmosphere around cooler host stars, in the same way they are in Venus’s atmosphere\cite{Jordan2021}. This means that with high precision retrievals of the upper atmosphere, the aerial biosphere hypothesis may be possible to test on Venus-like exoplanets. The framework that we present in this paper, self-consistently coupling metabolic reaction networks with a photochemical-kinetics network for the atmosphere of Venus, is equally applicable to simulating the effect of an aerial biosphere on the atmospheric chemistry of potentially habitable Venus-like exoplanets. If the new and exotic biome of a strictly aerial biosphere is possible in principle, then given the sheer number and diversity of exoplanets, we might expect that the first detected sign of life beyond the Solar System will originate from an aerial biosphere harboured on a world interior to the surface-water habitable zone.

\appendix

\section*{Methods}
\label{sec:methods}

\subsection*{The model.} We model the atmosphere of Venus using a photochemical-diffusion model \cite{Rimmer2021}. The model is composed of a $1{\rm D}$ Lagrangian solver, \textsc{Argo}, and a network of reactions, \textsc{Stand2020}, that treats H/C/N/O/S/Cl chemistry accurately within a temperature range of $100 - 30,000\,{\rm K}$ \cite{RimmerHelling2016,Hobbs2021}. The model has been shown to reproduce the known atmosphere of Venus within approximately an order of magnitude of observations given the inclusion of a scheme of aqueous chemistry that aids the gas-phase depletion of \ce{SO2} and \ce{H2O} \cite{Rimmer2021}. The chemical network, \textsc{Stand2020}, is a list of reactants, products, and rate constants, for every chemical reaction considered in the atmosphere. For thermochemical reactions the reaction rates are temperature dependent, and for photochemical reactions the reaction rates are dependent on the photon flux as a function of wavelength incident at a given layer of the atmosphere. These chemical reactions determine the chemical production and loss rates of each species. At each altitude step in the atmosphere the reactions are solved by \textsc{Argo} as a set of time-dependent, coupled, non-linear differential equations:

\begin{equation}
\dfrac{dn_{\rm X}}{dt} = P_{\rm X} - L_{\rm X}n_{\rm X} - \dfrac{\partial \Phi_{\rm X}}{\partial z},
\label{eq:argo_eqn}
\end{equation}
where, at a given height $z\,{\rm (cm)}$ and time $t\,{\rm (s)}$, $n_{\ce{X}}$ (cm$^{-3}$) is the number density of species $\ce{X}$, $P_{\rm X}$ (cm$^{-3}$ s$^{-1}$) is the rate of production of species \ce{X}, $L_{\rm X}$ (s$^{-1}$) is the rate constant for loss of species \ce{X}, and $\partial \Phi_{\rm X}/\partial z$ (cm$^{-3}$ s$^{-1}$) describes the divergence of the vertical diffusion flux, encapsulating both eddy- and molecular-diffusion.

\textsc{Argo} follows a parcel of gas as it rises from the surface to the top of the atmosphere and back down again. An initial condition for the chemical composition is input at the base of the atmosphere, listed in table \ref{tab:init}. At every altitude step on the journey upwards \textsc{Argo} solves equation \ref{eq:argo_eqn} for all species in the atmosphere, based on the pressure, temperature, and species' abundances at that altitude in the atmosphere. The time interval over which \textsc{Argo} solves for is prescribed by the eddy diffusion profile which parametrises vertical transport through the atmosphere. The pressure-temperature profile and eddy-diffusion profile for modern Venus are each taken from past photochemical-kinetics models of the lower \cite{Kras2007} and middle  \cite{Kras2012} atmosphere. At the top of the atmosphere, \textsc{Argo} takes the incident stellar spectrum and, at each step on the journey downwards, includes photochemical reactions driven by the stellar flux irradiating that layer of the atmosphere. \textsc{Argo} iterates this procedure until a convergence criteria is met and thus a global solution is found.
\begin{table}[h!]
\centering
\begin{tabular}{ c c c c c c c c c }
\hline\hline
\ce{CO_2} & \ce{N_2} & \ce{SO_2} & \ce{H_2O} & \ce{CO} & \ce{OCS} & \ce{HCl} & \ce{H_2} & \ce{H_2S} \\
\hline
0.96 & 0.03 & 150 ppm & 30 ppm & 20 ppm* & 5 ppm & 500 ppb & 3 ppb* & 10 ppb* \\
\hline
\end{tabular}
\caption{Initial surface abundances for the fiducial atmosphere \cite{Rimmer2021}. *The initial surface abundance of species marked with an asterisk are varied for some of the models. \label{tab:init}}
\end{table}

We test the effect of each of the net metabolic reactions, \ref{eq:phot_reaction}, \ref{eq:chem_1_reaction} and \ref{eq:chem_2_reaction} in turn, by coupling them to our atmospheric network and solving for the atmospheric composition of Venus. For each proposed metabolism the individual net metabolic reaction is added to the \textsc{Stand2020} reaction network with an effective rate constant that is unconstrained. This rate is not synonymous with the biochemical rate of the metabolic reaction itself, rather it is an `effective' net volumetric rate of all microbial activity in our one dimensional and diurnally averaged model. In this sense, the effective rate of the reaction relates to the biomass density in the model. The effective reaction is constrained to occur within the temperate region of the cloud layer by prescribing a condition that depends on atmospheric temperature. For a given effective rate, $k$, the reaction constant in the network depends on temperature (and thus altitude / pressure) as:
\begin{equation}
  {\rm Rate} =\begin{cases}
    k, & \text{if $273 < T\,{\rm (K)} < 373$}.\\
    0, & \text{otherwise}.
  \end{cases}
  \label{eq:rate_temp_condition}
\end{equation}

With each metabolism coupled to the \textsc{Stand2020} network in turn, we iterate \textsc{Argo} over a grid of models to explore the parameter space of net metabolic rate versus reagent abundance. The \ce{SO2} abundance remains fixed at it's fiducial value. For metabolism A we gradually increase the \ce{H2S} abundance from $10\,{\rm ppb}$ up to $500\,{\rm ppm}$. For metabolism B we gradually increase the \ce{H2} abundance from $3\,{\rm ppb}$ up to $10\,{\rm ppm}$, and then increase both the \ce{H2} and \ce{CO} abundances in tandem in the ratio 2:1 respectively, from $10\,{\rm ppm}$ up to $300\,{\rm ppm}$ for \ce{H2} and from $20\,{\rm ppm}$ up to $600\,{\rm ppm}$ for \ce{CO}. For metabolism C we gradually increase the \ce{CO} abundance from $20\,{\rm ppm}$ up to $650\,{\rm ppm}$.

\subsection*{Biomass density calculation.} For the model atmospheres that incorporate the metabolic activity of a putative Venusian biosphere at the maximum possible limit before transgressing observational constraints, we can estimate the maximum biomass density averaged over the volume of the cloud layer, under the assumption that the availability of metabolic inputs is the limiting factor on the productivity of the biosphere (in practise other factors, such as availability of metals for catalytic enzymes, will likely limit the productivity of the biosphere). For volumetric reaction rate $R$ ($\rm s^{-1}\,m^{-3}$) of the metabolic reaction in the model, we convert to the rate of energy liberated metabolically per unit volume in the cloud layer, $\epsilon$ ($\rm W\,m^{-3}$):
\begin{equation}
    \epsilon = \frac{R \, \Delta G_{\rm r}}{N_{\rm A}},
    \label{eq:energy}
\end{equation}
where $\Delta G_{r}$ ($\rm J\,mol^{-1}$) is the Gibbs free energy release from the metabolic reaction and $N_{\rm A}$ is Avogadro's constant. The average cell number density in the cloud layer, $n_{\rm cell}$ ($\rm cells\,m^{-3}$), is then given by:
\begin{equation}
    n_{\rm cell} = \frac{\epsilon}{P_{\rm cell}},
    \label{eq:n_cell}
\end{equation}
and the average biomass density:
\begin{equation}
    m_{\rm bio} = \frac{\epsilon\,M_{\rm cell}}{P_{\rm cell}},
    \label{eq:m_bio}
\end{equation}
where $P_{\rm cell}$ ($\rm W\,cell^{-1}$) represents the power requirement of a single microbe of mass $M_{\rm cell}$ (mg). This allows for a comparison to be made to the measured mass loading of aerosol in the atmosphere, and to the estimated mass loading of microbes in the Earth's (transiently) aerial biosphere. We can further make comparison to measured cell concentrations in analogue terrestrial environments that are most similar to the hyperarid environment of the Venusian cloud layer. To do so, we estimate the volume of aerosol per unit volume of the atmosphere, $v_{\rm aero}$ ($\rm mL\,m^{-3}$), within the cloud layer, from the reported aerosol mass loading $m_{\rm aero}$ ($\rm mg\,m^{-3}$) and an assumed density of aerosol material, $\rho_{\rm aero}$ ($\rm g\,mL^{-1}$):
\begin{equation}
    v_{\rm aero} = \frac{m_{\rm aero}}{\rho_{\rm aero}}.
    \label{eq:v_aero}
\end{equation}
The cell concentration within the aerosol substrate, $n_{\rm cell,\,aero}$ ($\rm cells\,mL^{-1}$), is then given by:
\begin{equation}
    n_{\rm cell,\,aero} = \frac{n_{\rm cell}}{v_{\rm aero}}.
    \label{eq:n_cell_aero}
\end{equation}
Combining equations (\ref{eq:energy} - \ref{eq:n_cell_aero}) and evaluating when the volumetric metabolic rate and metabolic input abundances are at the limit of transgressing observational data, gives us our final set of equations used to evaluate the range of maximum possible biomass densities and maximum possible cell concentrations referenced in the main text:

\begin{align}
    m_{\rm bio,\,max} &= \frac{R\,\Delta G_{\rm r}\,M_{\rm cell}}{N_{\rm A}\,P_{\rm cell}},
    \label{eq:m_bio_max} \\
    n_{\rm cell,\,aero,\,max} &= \frac{R\,\Delta G_{\rm r}\,\rho_{\rm aero}}{N_{\rm A}\,m_{\rm aero}\,P_{\rm cell}}.
    \label{eq:n_cell_aero_max}
\end{align}
The analysis here refers primarily to the chemotrophic metabolisms where $\Delta G_r \sim 240\,{\rm kJ\,mol^{-1}}$ is known a priori. Photosynthetic life would obtain energy from sunlight rather than a redox disequilibrium in the atmosphere, however the phototrophic metabolism can be treated with the same analysis if an assumption is made about the wavelength of light absorbed in reaction \ref{eq:photosystem}. For example, the energy delivered in sunlight at a wavelength of $500\,{\rm nm}$ is approximately $\sim 240\,{\rm kJ\,mol^{-1}}$ (i.e. per ${\rm N_A}$ of photons absorbed where ${\rm N_A}$ is Avogadro's constant) and so the above analysis can be repeated exactly for the phototroph (metabolism A) give or take some efficiency factor, and some numerical factor for the different possible wavelengths of light being utilised (the exception to this energetic analysis would be phototrophic life that captures light energy for cell maintenance without any redox chemistry occuring, in which case the potential biomass cannot be constrained by the rate of consumption of metabolic input species). Recent radiative transfer modelling has shown that, below $\sim 59\,{\rm km}$ altitude, while biologically-damaging UV radiation is screened by the cloud layer, photosynthetically active light at longer wavelengths ($>400\,{\rm nm}$) continues to penetrate \cite{Patel2022}. We therefore assume $\Delta G_r \sim 240\,{\rm kJ\,mol^{-1}}$ for each of the three net metabolic reactions. While our choice of $500\,{\rm nm}$ for the wavelength utilised in the photosynthetic metabolism is for simplicity in calculating the biomass, this value sits well within the range of common terrestrial photosynthetic wavelengths ($\sim 400 - 1000\,{\rm nm}$) including the wavelengths utilised by anoxygenic phototrophs in deep water columns \cite{Haas2018}. We further assume a typical value for cell density of a microbe to be of the order $\rho_{\rm cell} \sim 1\,{\rm g\,cm^{-3}}$ and typical size of $r_{\rm cell} \sim 0.5\,{\rm \mu m}$ in order to evaluate $M_{\rm cell}$. We can then investigate how the choice of cell power requirement, $P_{\rm cell}$, affects our estimates for the biomass density, in figure \ref{fig:biomass}. 

The required introduction of the cell power requirement term provides the largest uncertainty to this energetic analysis, varying over at least 10 orders of magnitude for terrestrial microbial life alone. It is for this reason why we must evaluate a range of possible estimates for $m_{\rm bio,\,max}$ and $n_{\rm cell,\,aero,\,max}$. Assuming that the stresses posed by the acidic and hyperarid Venusian environment pose greater survival challenges to life compared to conditions encountered on Earth, we can significantly constrain our estimated range of values of $m_{\rm bio}$ and $n_{\rm cell,\,aero}$ at the limiting metabolic reaction rate, by imposing that $P_{\rm cell} \gtrsim P_{\rm max,\,\bigoplus}$:

\begin{align}
    m_{\rm bio,\,max} &\lesssim \frac{R\,\Delta G_{\rm r}\,M_{\rm cell}}{N_{\rm A}\,P_{\rm max,\,\bigoplus}},
    \label{eq:bio_density}\\
    n_{\rm cell,\,aero,\,max} &\lesssim \frac{R\,\Delta G_{\rm r}\,\rho_{\rm aero}}{N_{\rm A}\,m_{\rm aero}\,P_{\rm max,\,\bigoplus}},
    \label{eq:bio_density}
\end{align}
Taking $P_{\rm max,\,\bigoplus}\,=\,4700 \times 10^{-15}\,{\rm J\,s^{-1}\,cell^{-1}}$ we evaluate limits $m_{\rm bio,\,max} \lesssim 3 \times 10^{-5}\,{\rm mg\,m^{-3}}$ for metabolism A, $m_{\rm bio,\,max} \lesssim 6 \times 10^{-4}\,{\rm mg\,m^{-3}}$ for metabolism B, and $m_{\rm bio,\,max} \lesssim 1 \times 10^{-3}\,{\rm mg\,m^{-3}}$ for metabolism C, as quoted in the abstract.

\section*{Data Availability}

The data generated in this study have been deposited in the Harvard online database under accession code (https://doi.org/10.7910/DVN/AKUTME).

\section*{Code Availability}

The methods underlying the code and the chemical network are all publicly available in the set of papers \cite{RimmerHelling2016,RimmerRugheimer2019,Hobbs2021,Rimmer2021}. The ARGO software may be made available upon reasonable request from the authors.

\section*{Acknowledgements}

S.J. thanks the Science and Technology Facilities Council (STFC) for the PhD studentship (grant reference ST/V50659X/1). P.B.R. thanks the Simons Foundation for funding (SCOL awards 599634).

\section*{Author Contributions}

S.J., O.S., and P.B.R. conceived of the study and contributed extensively to the manuscript. S.J. implemented the simulations and analysed the output data. 

\section*{Competing Interests}

The authors declare no competing interests.

\end{document}